# Fast Hebbian plasticity and working memory

Anders Lansner[a,b] (ala@kth.se), Florian Fiebig[b] (fiebig@kth.se), Pawel Herman[b,c] (paherman@kth.se)
[a]Stockholm University, Department of Mathematics,SE-106 91 Stockholm, Sweden
[b]KTH Royal Institute of Technology, Dept of Computational Science and Technology, 100 44 Stockholm, Sweden
[c]Digital Futures, KTH Royal Institute of Technology

Corresponding author: Anders Lansner

*Abstract*

Theories and models of working memory (WM) were at least since the mid-1990s dominated by the persistent activity hypothesis. The past decade has seen rising concerns about the shortcomings of sustained activity as the mechanism for short-term maintenance of WM information in the light of accumulating experimental evidence for so-called activity-silent WM and the fundamental difficulty in explaining robust multi-item WM. In consequence, alternative theories are now explored mostly in the direction of fast synaptic plasticity as the underlying mechanism.The question of non-Hebbian vs Hebbian synaptic plasticity emerges naturally  in this context. In this review we focus on fast Hebbian plasticity and trace the origins of WM theories and models building on this form of associative learning.

## Introduction

A tight interaction between different memory systems is critical for most functions of the human brain, not least its cognitive capabilities. There is a general consensus in the brain science community regarding synaptic plasticity as the most important neural mechanism behind different forms of semantic and episodic long-term memory (LTM). However, there is less agreement about mechanisms underlying short-term memory (STM) phenomena, and in particular working memory (WM), which supports temporary maintenance of information as well as flexible manipulation so that the brain can keep record of the situation at hand and retain a coherent view of the here and now [1,2]. Although WM has a memory capacity of only a handful of items it can activate associated long-term memory representations and thereby orchestrate perceptual and cognitive processes. The prefrontal cortex (PFC) is considered a critical brain region for WM and reciprocal interactions with hippocampus and neocortical memory stores are prominent [3].

Hebbian associative synaptic plasticity in the form of long-term potentiation (LTP) is today well established in neuroscience as the primary neural mechanism behind long-term memory [4]. In contrast, WM was for long regarded as solely dependent on persistent activity in recurrent PFC circuits [5], commonly modelled as attractor neural networks. However, recently the original persistent activity WM hypothesis has been challenged and different forms of fast synaptic plasticity proposed to support WM. The emerging synaptic theories and models have potential to unify computational principles behind memory phenomena in the brain across different timescales, from WM to LTM. Despite and partly because of the recent focus in the literature on non-Hebbian forms of synaptic plasticity as a key WM mechanism alternative to persistent activity, we advocate the potential of fast Hebbian plasticity for supporting robust WM. In particular, we review the underlying computational theories, functional implications and experimental evidence for the involvement of fast inducing and expressing associative ("Hebbian") synaptic plasticity in WM.

## From persistent activity to synaptic WM

An intense debate about the mechanistic underpinnings of WM phenomena has been largely fuelled by experimental accounts of neural activity during the so-called delay period, considered to be an operational hallmark of WM. Pioneering work [6,7] laid the foundation for the so-called persistent activity hypothesis [5], postulating that sustained activity of neuronal populations throughout the delay period is the key WM mechanism. Subsequently, there have been many hypotheses proposed to explain the neurobiological basis of persistent activity in the cortex, ranging from the intrinsic properties of individual neurons to the connectivity and properties of neural circuits [8–10]. Computational efforts aimed at capturing the prolonged sustained neuronal firing effects focused on attractor dynamics in recurrent networks, as in, for example, the so-called bump attractor model of an oculomotor delayed-response WM experiment [11,12].

With a growing body of theoretical and experimental work the concerns about limitations of this original theory and its computational modelling implementations have accumulated. The early bump models are inherently single-item memories as they cannot reliably store multiple attractor memories at the same time or even alternate between them, which is necessary for more advanced forms of WM [13]. Other distributed attractor memory networks can store a number of items in their weight matrix and activate them spontaneously or in response to a stimulus [14]. In experiments involving multi-item WM tasks, bursting rather than sustained activity has been reported as a correlate of WM maintenance and control, much as these computational models demonstrate [15]. Furthermore, in a striking contrast to the robustness of biological WM and distributed attractor memory networks, the bump models are typically vulnerable to distractors or other sources of interfering activity unless additional stabilising factors are included [12]. They also have problems reproducing activity-silent memory, which suggests that WM information can be robustly maintained without the need for neural populations to be persistently active [16].

Ultimately, in the light of accumulating experimental evidence demonstrating that WM information can be sustained without persistent activity there is a growing appreciation for alternative views on the underlying neural mechanisms. A deepening synergy between experimental and computational studies supports a steady paradigm shift. In fact, some early theories and computational models proposed important roles for Hebbian synaptic plasticity in sensory fusion, neural binding, and WM [17–19], displaying periodic oscillations and reactivations or time-multiplexing of different internal representations and thus a transient nature of delay activity. However, they received limited attention, presumably due to a prevailing view that induction of LTP-like plasticity needs repetitive stimuli and is slow to express together with the very sparse biological evidence for such fast forms of Hebbian synaptic plasticity. Therefore, before we discuss Hebbian synaptic WM network models in more detail, we briefly review the experimental data regarding these processes.

## Experimental data on fast Hebbian synaptic plasticity

Synaptic plasticity is a highly diverse and complex phenomenon with many different aspects and components [20]. Yet, in the form of Hebbian long-term potentiation (LTP) it is well established as a key mechanism behind LTM. There is much less experimental data on fast Hebbian synaptic plasticity in cortex, but it was nevertheless described in the 1990's as one component of the early LTP, referred to as short-term potentiation (STP) [21,22]. Induction was possible with 5 to 10 spikes within a theta cycle and expression could happen within 2-3 seconds [23]. Later studies emphasised the possible role of such Hebbian-STP in WM [24,25]. It was suggested that the underlying molecular mechanism is NMDA activation by glutamate causing $Ca^{2+}$ inflow and triggering phosphorylation and increased conductance of AMPA receptor gated ion channels [25]. More recently, different forms of Hebbian-STP have been described and some of these display a use-dependent decay [26,27], quite appropriate for a palimpsest type of memory, which can avoid catastrophic forgetting [28]. Unfortunately, fast synaptic plasticity is quite difficult to study in detail with current experimental technology, in part because LTP induction protocols and baseline monitoring may disrupt STP [26], but all-optical technology for targeted stimulation and voltage recording in vivo may change that [29].

A more indirect indication of Hebbian-STP-like properties of WM is that blocking NMDA channels with ketamine critically affects WM function [30] (but see [31] for an alternative explanation). Dopamine inputs to the PFC play a prominent role in normal cognitive processes and neuropsychiatric pathologies [32,33]. In analogy with its action in basal ganglia [34], dopamine may also be involved in synaptic plasticity in the PFC [35,36].

# Computational models of Hebbian synaptic WM

The experimental data on fast Hebbian synaptic plasticity and evidence for its involvement in WM is still quite limited. Despite this, data-constrained computational models have contributed to an improved understanding of the functional consequences for WM and provided experimentally testable predictions. A key component of these network models is their synaptic plasticity mechanisms so we first give a brief account of these.

## Models of synaptic plasticity

More abstract network simulations with non-spiking units often use standard correlation based learning rules, see e.g. [37]. Here we focus on somewhat more detailed plasticity models adapted for spiking neural network models (SNN). We do not cover computational models that represent biochemical and biophysical synaptic processes here, though these are relevant in the domain of neuropharmacology [38,39].

*Spike-timing dependent plasticity (STDP)*

STDP is by far the most commonly used synaptic plasticity model for SNN. It is a data driven, phenomenological model [40,41] in which the weight change is dependent on the relative timing of pre- and postsynaptic spikes. A wide variety of symmetric and asymmetric STDP kernels have been described and evaluated, and a triplet-STDP model was developed as an extension to achieve a better fit with experimental data [42,43]. A reward-dependent STDP kernel with eligibility traces, r-STDP, was developed by Izhiekevich [44] and employed to model synaptic WM [45]. Notably, STDP models can be related to non-spiking rate models under the assumption of Poisson firing of both pre- and postsynaptic neurons and can then reproduce standard rate-based Hebbian as well as anti-Hebbian learning [46].

*The Bayesian Confidence Propagation learning rule (BCPNN)*

The BCPNN model has been used in several recent models of Hebbian synaptic WM. Its learning rule was derived in a straightforward way from Bayes rule, first for non-spiking models of cortical associative memory but later also adapted for SNN [47–49]. Activation and co-activation probabilities are estimated by exponentially weighted moving averages, including an eligibility trace. The learning rate is determined by the time constant of the slowest moving of these - high values produce long-term and low values short-term memory [48]. Pre-post activity correlation over time produces an excitatory synapse whereas lack of correlation gives a zero connection, and anti-correlation generates inhibition, mapped to dendritic targeting di-synaptic inhibition [50]. Further, unlike most Hebbian plasticity models, BCPNN adjusts neuronal intrinsic excitability of the postsynaptic neuron to reflect its activity history, corresponding to a Bayesian prior [49,51].

## Cortical WM network models

The earliest model explicitly employing fast Hebbian synaptic plasticity for WM [19] focused on explaining delay activity. Unlike the early bump attractor models [12], it could perform one-shot encoding of multiple items and displayed features of activity-silent WM, similar to later proposed models based on non-Hebbian-STP [52]. However, the Hebbian mechanism made the network robust against distractors and network inhomogeneity. A variant of this Hebbian-STP based model was relying on a removed NMDA Mg-block for synaptic potentiation [53], an idea revived later by Szatmary and Izhikevich [54] using a somewhat unique hybrid model also featuring embedded LTM structures (polychronous groups), formed by LTP. One-shot formation of long-lived novel memories via fast STDP was also studied in elaborate spiking network models [55,56].

The most recent Hebbian WM models implement rate-based covariance rules [57,58], or more detailed spiking learning rules, such as reward based learning to store task relevant WM representations or BCPNN [45,59,60].

*Beyond delay activity*

Attractor WM network models with fast Hebbian synaptic plasticity show more or less sustained delay activity during memory readout, similar to other models with fixed connectivity and non-Hebbian-STP mechanisms. Together with the latter, they also display multi-item and activity-silent WM. However, the somewhat narrow focus on maintenance in the past has overemphasised such mechanisms whilst overlooking other important aspects of WM function, some of which we address in the following.

---

**Box 1. Hebbian but not non-Hebbian plasticity can encode novel memories**

A. Forming novel memories and associations is a hallmark of human memory. The figure below illustrates that the outcome of non-Hebbian-STP is unspecific by enhancing all outgoing synapses from neurons activated by a stimulus. In contrast, Hebbian plasticity potentiates selectively synapses targeting only active postsynaptic neurons and thus forms a retrievable memory. As a result, cue-driven recall in neural circuits with only non-Hebbian-STP leads to a spreading activation, unlike in circuits with Hebbian-STP, where a retrievable memory is encoded.

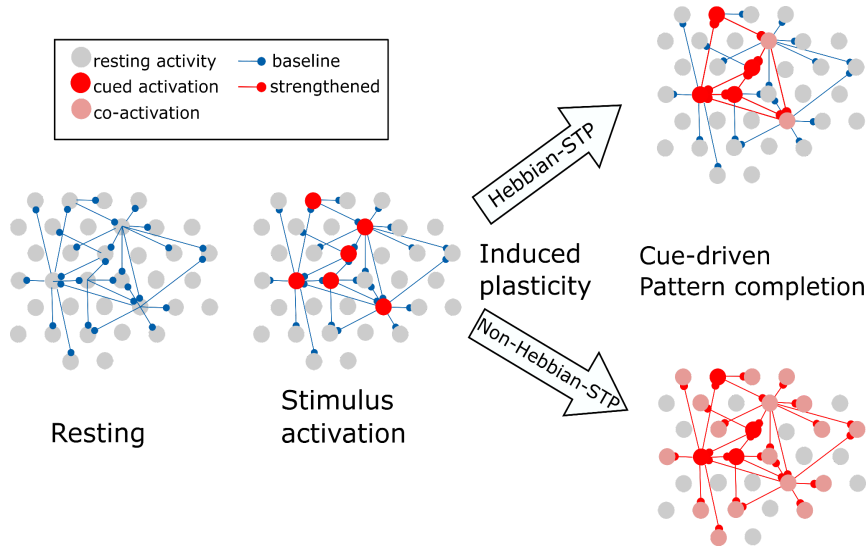

B. The low discriminability of non-Hebbian plasticity is further illustrated by the below weight matrices resulting from encoding seven sparse random patterns in a small recurrent attractor memory using Hebbian compared to pre- and postsynaptic non-Hebbian plasticity. Synaptic facilitation or augmentation are examples of presynaptic non-Hebbian-STP and an intrinsic excitability change is an example of postsynaptic non-Hebbian-STP.

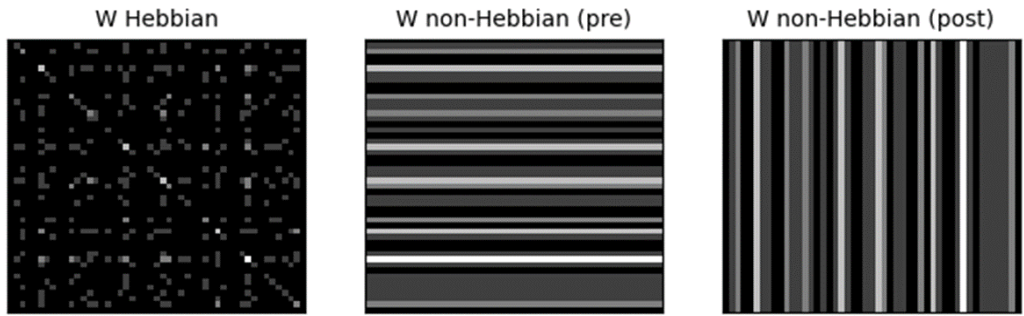

***Novel encoding.*** Fast Hebbian changes in synaptic strength permits the formation of WM representations from initially unstructured circuits, and several authors have stressed this aspect of WM (see e.g. [19,53,54,59]). It is important to realise that novel item learning (or the learning of novel conjunctions of already stored items) cannot be achieved by presynaptic mechanisms alone (see Box 1). Non-Hebbian models typically build on pre-existing memory representations. Manohar et al. [57] replaced their Hebbian learning component with non-Hebbian-STP and demonstrated that the resulting model could not achieve above-chance encoding of novel WM items. On the other hand, we have recently demonstrated that one-shot encoding of attractor memories is more potent in a model that synergistically exploits both non-Hebbian-STP and Hebbian BCPNN plasticity ( Fig. 1a, unpublished data)

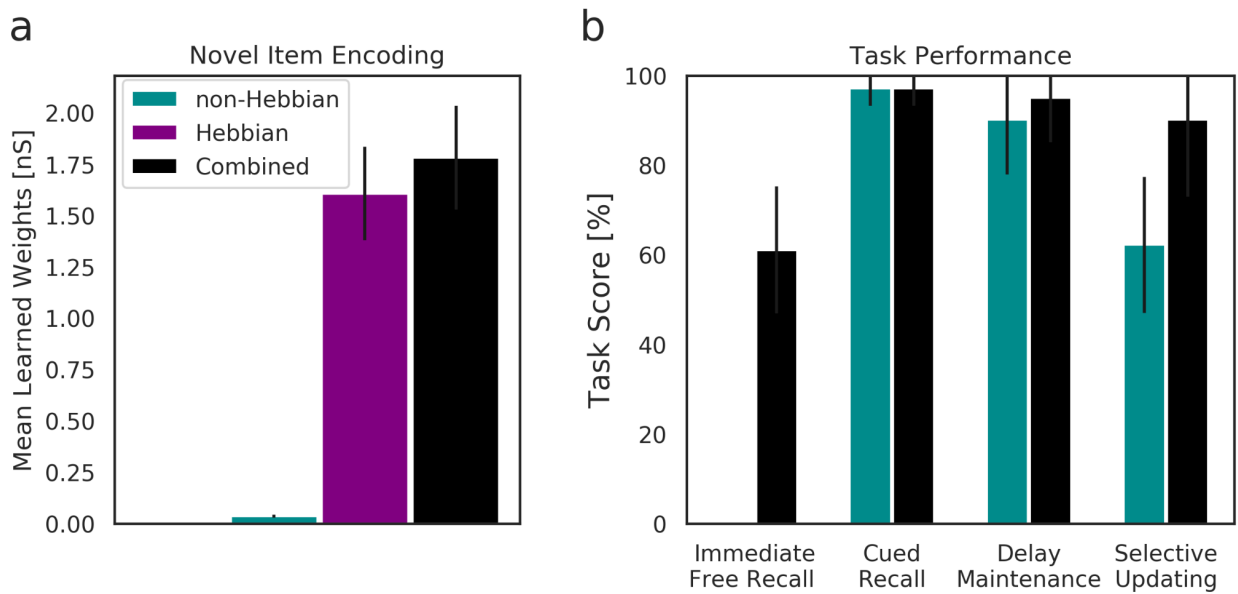

Figure 1. WM Encoding, Maintenance, and Updating in an SNN with Hebbian, non-Hebbian, or combined synaptic plasticity, based on an unpublished extension of the model by Fiebig & Lansner (2017). a) Mean recurrent weights between pyramidals following novel item learning. b) Task performance of model configurations in immediate free recall of novel items, cued recall, delay maintenance of four cued items over five seconds, and replacement (selective updating) of the maintained items with four new items.

***Selective updating and reset***. In daily WM scenarios memory items need to be efficiently updated or removed from WM, depending on dynamically evolving task demands. This includes the possibility of wiping out the entire WM by a burst of uncorrelated noise that resets Hebbian weights [19]. Hebbian learning typically induces memories with palimpsest properties (see e.g. [28]), so that items are not removed by time but by additional stimuli. Mechanisms that incorporate inhibitory learning or multiplex the learning of excitation and inhibition (e.g. BCPNN) can extend that capability even further, by replacing a set of maintained memories with a new set of memories [60].

***STM - LTM indexing.*** WM is multi-modal and compositional so it can robustly leverage existing long-term memory [2,61] and sensory representations [9]. Hebbian models can perform this function by momentarily binding long-term memories or feature selective populations through a dynamically recruited STM population, putatively in PFC. In such WM models, the recurrent STM population is essentially a temporary cell assembly [57,60], which rapidly forms a joint distributed attractor with the sensory LTM representation via fast Hebbian STM–>LTM plasticity. This connection thereafter acts as an index to the bound LTM representations. It has been pointed out that such a distributed attractor network could without any modification perform visual search [58] and variable-value binding [62] (Fig. 2). It can further explain observations of dynamic WM representations [63], since the dynamic PFC cell assemblies may disappear and rebuild during an ongoing task [36].

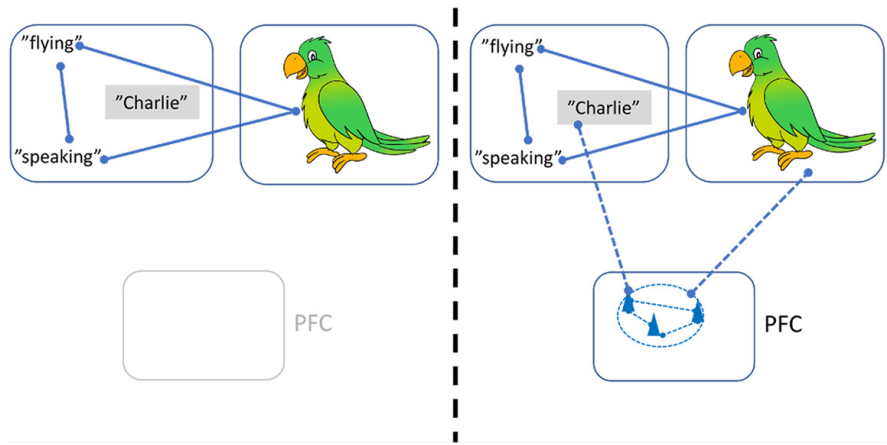

Figure 2. Variable-value binding could be achieved by PFC short-term STM-LTM indexing. Left: Initially, the multimodal representation of "parrot" exists in LTM side by side with but disconnected from "Charlie" as a representation of a proper name. When someone states that "Charlie is my parrot," the name "Charlie" is temporarily and reciprocally bound to the parrot representation via PFC, mediated by fast Hebbian plasticity. Right: Pattern completion now allows "Charlie" to trigger the entire assembly and, analogously, makes "flying" or the sight of a parrot trigger "Charlie."

## Concluding remarks

There is a tangible paradigm shift in the discussion about the fundamental mechanisms underlying WM. While persistent activity may still be observed in specific WM tasks - particularly as a correlate of memory activation - synaptic plasticity is now clearly in focus. In this review we have concentrated on a particularly promising hypothesis that Hebbian-STP, akin to LTP but with a faster induction and expression, is a key mechanism behind WM. We acknowledge however that the biological evidence is not yet conclusive. There is a growing expectation that new experimental techniques offering sufficient spatial and temporal resolution in a synergistic combination with computational models will finally reveal deeper insights into the correlates and mechanisms of different aspects of WM function, ranging from encoding, maintenance, updating to volitional control.

## Author contributions

All authors discussed and wrote the paper.

## Conflicts of interest

Nothing to declare.

## Acknowledgements

We would like to thank the Swedish Research Council (VR) grants: 2018-05360 and 2016-05871, Digital Futures and Swedish e-science Research Center (SeRC) for their support.

# References

1. D'Esposito M, Postle BR. The cognitive neuroscience of working memory. Annu Rev Psychol. 2015;66: 115–142.

2. Eriksson J, Vogel EK, Lansner A, Bergström F, Nyberg L. Neurocognitive Architecture of Working Memory. Neuron. 2015. pp. 33–46. doi:10.1016/j.neuron.2015.09.020

3. Myroshnychenko M, Seamans JK, Phillips AG, Lapish CC. Temporal Dynamics of Hippocampal and Medial Prefrontal Cortex Interactions During the Delay Period of a Working Memory-Guided Foraging Task. Cerebral Cortex. 2017. pp. 5331–5342. doi:10.1093/cercor/bhx184

4. Langille JJ, Brown RE. The Synaptic Theory of Memory: A Historical Survey and Reconciliation of Recent Opposition. Front Syst Neurosci. 2018;12: 52.

5. Goldman-Rakic PS. Architecture of the prefrontal cortex and the central executive. Ann N Y Acad Sci. 1995;769: 71–83.

6. Fuster JM, Alexander GE. Neuron activity related to short-term memory. Science. 1971;173: 652–654.

7. Kubota K, Niki H. Prefrontal cortical unit activity and delayed alternation performance in monkeys. Journal of Neurophysiology. 1971. pp. 337–347. doi:10.1152/jn.1971.34.3.337

8. Zylberberg J, Strowbridge BW. Mechanisms of Persistent Activity in Cortical Circuits: Possible Neural Substrates for Working Memory. Annu Rev Neurosci. 2017;40: 603–627.

9. Sreenivasan KK, Curtis CE, D'Esposito M. Revisiting the role of persistent neural activity during working memory. Trends Cogn Sci. 2014;18: 82–89.

10. Durstewitz D, Seamans JK, Sejnowski TJ. Neurocomputational models of working memory. Nat Neurosci. 2000;3 Suppl: 1184–1191.

11. Camperi M, Wang XJ. A model of visuospatial working memory in prefrontal cortex: recurrent network and cellular bistability. J Comput Neurosci. 1998;5: 383–405.

12. Compte A, Brunel N, Goldman-Rakic PS, Wang X-J. Synaptic Mechanisms and Network Dynamics Underlying Spatial Working Memory in a Cortical Network Model. Cereb Cortex. 2000;10: 910–923.

13. Miller P. Itinerancy between attractor states in neural systems. Curr Opin Neurobiol. 2016;40: 14–22.

14. Amit DJ, Brunel N. Model of global spontaneous activity and local structured activity during delay periods in the cerebral cortex. Cereb Cortex. 1997;7: 237–252.

15. Lundqvist M, Rose J, Herman P, Brincat SL, Buschman TJ, Miller EK. Gamma and Beta Bursts Underlie Working Memory. Neuron. 2016. pp. 152–164. doi:10.1016/j.neuron.2016.02.028

16. ** Stokes MG. "Activity-silent" working memory in prefrontal cortex: a dynamic coding framework. Trends Cogn Sci. 2015;19: 394–405.

**A seminal review of other than persistent activity correlates of WM in prefrontal cortex, highlighting the importance of neural dynamics for WM. It discusses possible neural substrates of non-Hebbian or Hebbian forms of STP as basis for so-called activity-silent WM.**

17. von der Malsburg C. Binding in models of perception and brain function. Curr Opin Neurobiol. 1995;5: 520–526.

18. Tononi G, Sporns O, Edelman GM. A measure for brain complexity: relating functional segregation and integration in the nervous system. Proceedings of the National Academy of Sciences. 1994. pp. 5033–5037. doi:10.1073/pnas.91.11.5033

19. ** Sandberg A, Tegnér J, Lansner A. A working memory model based on fast Hebbian learning. Network: Computation in Neural Systems. 2003. pp. 789–802. doi:10.1088/0954-898X/14/4/309.
**The first recurrent attractor WM model based on fast Hebbian synaptic plasticity. A modular network with adapting, non-spiking units and the BCPNN learning rule displays one-shot novel encoding, multi-item palimpsest STM and updating, high distractor tolerance, and spontaneous replay of stored memories.**

20. Sjöström PJ, Turrigiano GG, Nelson SB. Rate, timing, and cooperativity jointly determine cortical synaptic plasticity. Neuron. 2001;32: 1149–1164.

21. Malenka RC. Postsynaptic factors control the duration of synaptic enhancement in area CA1 of the hippocampus. Neuron. 1991;6: 53–60.

22. * Malenka RC, Nicoll RA. NMDA-receptor-dependent synaptic plasticity: multiple forms and mechanisms. Trends Neurosci. 1993;16: 521–527.
**An early extensive review of experimental findings about synaptic plasticity processes operating during the first 1-2 h of LTP. It highlights hippocampal NMDA-dependent Hebbian-STP, conditions for its induction and expression as well as relation to long-lasting LTP components.**

23. Gustafsson B, Asztely F, Hanse E, Wigstrom H. Onset Characteristics of Long-Term Potentiation in the Guinea-Pig Hippocampal CA1 Region in Vitro. European Journal of Neuroscience. 1989. pp. 382–394. doi:10.1111/j.1460-9568.1989.tb00803.x

24. ** Erickson MA, Maramara LA, Lisman J. A Single Brief Burst Induces GluR1-dependent Associative Short-term Potentiation: A Potential Mechanism for Short-term Memory. Journal of Cognitive Neuroscience. 2010. pp. 2530–2540. doi:10.1162/jocn.2009.21375.
**The earliest paper that proposes and treats in some depth fast Hebbian-STP as a key component in WM. Reviews information on induction expression and duration critical for WM function.**

25. Lisman J. Glutamatergic synapses are structurally and biochemically complex because of multiple plasticity processes: long-term potentiation, long-term depression, short-term potentiation and scaling. Philos Trans R Soc Lond B Biol Sci. 2017 [cited 11 Jan 2023]. doi:10.1098/rstb.2016.0260

26. Volianskis A, Bannister N, Collett VJ, Irvine MW, Monaghan DT, Fitzjohn SM, et al. Different NMDA receptor subtypes mediate induction of long-term potentiation and two forms of short-term potentiation at CA1 synapses in rat hippocampus in vitro. J Physiol. 2013;591: 955–972.

27. Volianskis A, Jensen MS. Transient and sustained types of long-term potentiation in the CA1 area of the rat hippocampus. J Physiol. 2003;550: 459–492.

28. Nadal JP, Toulouse G, Changeux JP, Dehaene S. Networks of Formal Neurons and Memory Palimpsests. Europhysics Letters (EPL). 1986. pp. 535–542. doi:10.1209/0295-5075/1/10/008

29. Fan LZ, Kim DK, Jennings JH, Tian H, Wang PY, Ramakrishnan C, et al. All-optical physiology resolves a synaptic basis for behavioral timescale plasticity. Cell. 2023. pp. 543–559.e19. doi:10.1016/j.cell.2022.12.035

30. Driesen NR, McCarthy G, Bhagwagar Z, Bloch MH, Calhoun VD, D'Souza DC, et al. The impact of NMDA receptor blockade on human working memory-related prefrontal function and connectivity. Neuropsychopharmacology. 2013;38: 2613–2622.

31. Wang M, Yang Y, Wang C-J, Gamo NJ, Jin LE, Mazer JA, et al. NMDA Receptors Subserve Persistent Neuronal Firing during Working Memory in Dorsolateral Prefrontal Cortex. Neuron. 2013. pp. 736–749. doi:10.1016/j.neuron.2012.12.032

32. Seamans JK, Yang CR. The principal features and mechanisms of dopamine modulation in the prefrontal cortex. Prog Neurobiol. 2004;74: 1–58.

33. Goto Y, Yang CR, Otani S. Functional and dysfunctional synaptic plasticity in prefrontal cortex: roles in psychiatric disorders. Biol Psychiatry. 2010;67: 199–207.

34. Morris G, Schmidt R, Bergman H. Striatal action-learning based on dopamine concentration. Exp Brain Res. 2010;200: 307–317.

35. Huang Y-Y, Simpson E, Kellendonk C, Kandel ER. Genetic evidence for the bidirectional modulation of synaptic plasticity in the prefrontal cortex by D1 receptors. Proc Natl Acad Sci U S A. 2004;101: 3236–3241.

36. ** Bocincova A, Buschman TJ, Stokes MG, Manohar SG. Neural signature of flexible coding in prefrontal cortex. Proc Natl Acad Sci U S A. 2022;119: e2200400119.
**Shows that characteristic trial-to-trial changes in neural selectivity generated by a WM model with Hebbian-STP are also observed in PFC neural populations of monkeys performing a WM task.**

37. Amit DJ, Mongillo G. Spike-driven synaptic dynamics generating working memory states. Neural Comput. 2003;15: 565–596.

38. Bhalla US. Molecular computation in neurons: a modeling perspective. Curr Opin Neurobiol. 2014;25: 31–37.

39. Kotaleski JH, Blackwell KT. Modelling the molecular mechanisms of synaptic plasticity using systems biology approaches. Nature Reviews Neuroscience. 2010. pp. 239–251. doi:10.1038/nrn2807

40. Bi GQ, Poo MM. Synaptic modifications in cultured hippocampal neurons: dependence on spike timing, synaptic strength, and postsynaptic cell type. J Neurosci. 1998;18: 10464–10472.

41. Song S, Miller KD, Abbott LF. Competitive Hebbian learning through spike-timing-dependent synaptic plasticity. Nat Neurosci. 2000;3: 919–926.

42. Morrison A, Diesmann M, Gerstner W. Phenomenological models of synaptic plasticity

based on spike timing. Biological Cybernetics. 2008. pp. 459–478. doi:10.1007/s00422-008-0233-1

43. Pfister J-P, Gerstner W. Triplets of spikes in a model of spike timing-dependent plasticity. J Neurosci. 2006;26: 9673–9682.

44. Izhikevich EM. Solving the distal reward problem through linkage of STDP and dopamine signaling. Cereb Cortex. 2007;17: 2443–2452.

45. Savin C, Triesch J. Emergence of task-dependent representations in working memory circuits. Front Comput Neurosci. 2014;8: 57.

46. Kempter R, Gerstner W, van Hemmen JL. Hebbian learning and spiking neurons. Physical Review E. 1999. pp. 4498–4514. doi:10.1103/physreve.59.4498

47. Lansner A, Holst A. A higher order Bayesian neural network with spiking units. Int J Neural Syst. 1996;7: 115–128.

48. Sandberg A, Lansner A, Petersson KM, Ekeberg O. A Bayesian attractor network with incremental learning. Network. 2002;13: 179–194.

49. Tully PJ, Hennig MH, Lansner A. Synaptic and nonsynaptic plasticity approximating probabilistic inference. Front Synaptic Neurosci. 2014;6. doi:10.3389/fnsyn.2014.00008

50. Chrysanthidis N, Fiebig F, Lansner A. Introducing double bouquet cells into a modular cortical associative memory model. Journal of Computational Neuroscience. 2019. pp. 223–230. doi:10.1007/s10827-019-00729-1

51. Debanne D, Inglebert Y, Russier M. Plasticity of intrinsic neuronal excitability. Curr Opin Neurobiol. 2019;54: 73–82.

52. Mongillo G, Barak O, Tsodyks M. Synaptic theory of working memory. Science. 2008;319: 1543–1546.

53. Lisman JE, Fellous J-M, Wang X-J. A role for NMDA-receptor channels in working memory. Nature Neuroscience. 1998. pp. 273–275. doi:10.1038/1086

54. * Szatmáry B, Izhikevich EM. Spike-timing theory of working memory. PLoS Comput Biol. 2010;6. doi:10.1371/journal.pcbi.1000879
**A recurrent network with polychronous groups formed by long-term STDP. Combines Hebbian-STP with NMDA plateau plasticity for multi-item WM. Novel memories can be encoded after repeated presentation.**

55. Del Giudice P, Fusi S, Mattia M. Modelling the formation of working memory with networks of integrate-and-fire neurons connected by plastic synapses. J Physiol Paris. 2003;97: 659–681.

56. Zenke F, Agnes EJ, Gerstner W. Diverse synaptic plasticity mechanisms orchestrated to form and retrieve memories in spiking neural networks. Nature Communications. 2015. doi:10.1038/ncomms7922

57. * Manohar SG, Zokaei N, Fallon SJ, Vogels TP, Husain M. Neural mechanisms of attending to items in working memory. Neurosci Biobehav Rev. 2019;101: 1–12.

**Describes how attention and working memory may interact. It demonstrates this in a small proof-of-concept non-spiking model with sensory and conjunctive neurons. Demonstrates how persistent activity and activity-silent maintenance processes could co-exist.**

58. Bocincova A, Olivers CNL, Stokes MG, Manohar SG. A common neural network architecture for visual search and working memory. Vis cogn. 2020;28: 356–371.

59. ** Fiebig F, Lansner A. A Spiking Working Memory Model Based on Hebbian Short-Term Potentiation. J Neurosci. 2017;37: 83–96.
**A spiking modular recurrent neural network model with Hebbian-STP encodes items from a list and freely recalls about 5-7 of those. The statistics of free and cued recall from the model reproduces results from human experiments in terms of primacy, recency and conditional recall probability.**

60. ** Fiebig F, Herman P, Lansner A. An Indexing Theory for Working Memory Based on Fast Hebbian Plasticity. eNeuro. 2020;7. doi:10.1523/ENEURO.0374-19.2020.
**Presents an indexing theory for and computational model of WM in macaque PFC and temporal cortex. The model features spiking units, cortical columns, laminar connection profiles, mixed receptor types, patchy long-range connectivity, and plausible connection delays. Temporary cell assemblies formed in PFC bind with neocortical LTM representations by fast Hebbian plasticity, thus activating LTM representations and providing novel encoding, maintenance, selective updating and dynamic coding in PFC.**

61. Fuster JM. Distributed memory for both short and long term. Neurobiol Learn Mem. 1998;70: 268–274.

62. Feldman J. The neural binding problem(s). Cogn Neurodyn. 2013;7: 1–11.

63. Stokes MG, Kusunoki M, Sigala N, Nili H, Gaffan D, Duncan J. Dynamic coding for cognitive control in prefrontal cortex. Neuron. 2013;78: 364–375.